# USING AN ONLINE LEARNING ENVIRONMENT TO TEACH AN UNDERGRADUATE STATISTICS COURSE: THE TUTOR-WEB

Anna Helga Jonsdottir[1], Gunnar Stefansson[1]

[1] *University of Iceland (Iceland)*
E-mails [ahj@hi.is, gunnar@hi.is]

## Abstract

A learning environment, the tutor-web (http://tutor-web.net), has been developed and used for educational research. The system is accessible and free to use for anyone having access to the Web. It is based on open source software and the teaching material is licensed under the Creative Commons Attribution-ShareAlike License. The system has been used for computer-assisted education in statistics and mathematics. It offers a unique way to structure and link together teaching material and includes interactive quizzes with the primary purpose of increasing learning rather than mere evaluation.

The system was used in a course on basic statistics in the University of Iceland, spring 2013. A randomized trial was conducted to investigate the difference in learning between students doing regular homework and students using the system. The difference between the groups was not found to be significant.

## 1 INTRODUCTION

The tutor-web is a web-based learning environment accessible and free to use for everyone having access to the Web. The teaching material in the system is linked together in a unique way making it easy for the student to browse through the material. The system also includes interactive quizzes with the primary purpose of increasing learning rather than mere evaluation. The system is solely based on open source software. It is written in Plone [1] which is a Web-based content management system (CMS) built on top of the Zope Application Server [2]. The teaching material is licensed under the Creative Commons Attribution-ShareAlike License [3] to provide usage of material to institutions of limited resources.

The tutor-web has been used in several courses at the University of Iceland. In the following analysis, data collected in a basic statistical course taught in 2013 will be used. It is of interest to measure possible difference in learning between students answering quiz questions in the tutor-web and students doing more traditional homework such as handing in written assignments. While no additional work is needed by teachers after tutor-web assignments have been handed in, written solutions needs to be corrected which can be very time consuming and therefore costly. The system could therefore save time and effort if it can replace traditional homework to some extent. To answer this question a randomized trial was conducted where students either worked within in the tutor-web as homework or handed in written assignments corrected by a teaching assistant.

A short summary of existing educational systems will be given in Section 2 and a brief introduction to the tutor-web in Section 3. The randomized trial is described in Section 4 along with the results of the analysis. Future work and summary of conclusions can be found in Section 5.

## 2 EXISTING EDUCATIONAL SYSTEMS

Several types of educational systems have emerged the past several years, these include learning management system (LMS), learning content management system (LCMS), virtual learning environment (VLE), course management system (CMS) and Adaptive and intelligent Web-based educational systems (AIWBES). The terms VLE and CMS are often used interchangeably, CMS being more common in the United States and VLE in Europe. Examples of LMS, LCMS, VLE, CMS or mixes thereof, include BlackBoard, Moodle, Atutor, Dokeos, ILIAS, OpenUSS, Clairoline, Sakai and the LON-CAPA system [4]. These can be used for administration of students and courses, creation and/or storing educational content, assessment and more. Classes taught on these platforms are accessible through a web-browser but are usually private, i.e. only individuals who are registered for the class have access to the password-protected website. Of these, Moodle is particularly widely used (being open-source).

An interesting feature of the LON-CAPA system are homework problems which are typically randomized in a way that each student will get an individual version of the problem to work on, differing for instance by the numerical values used in the problem. This technology, often called parameterized questions, allows the production of many questions from a small number of templates able to produce many similar, yet different questions. The use of this type of individualized exercises has been shown to significantly reduce cheating, while at the same time improving student understanding and exam performance [5]. Other systems using parameterized questions include WebAssign in math and science, QuizJet [6] in the Java Programming Language, the Mallard system [7] and QuizPACK [8] for programming-related courses language. A study of students using the QuizPACK system showed that successful work in the system correlated with the students success on classroom quizzes and final exam performance. The amount of student work with the parameterized questions was also found to be a significant predictor of their course knowledge gain [9].

Number of content providers can be found on the internet. Even though they are not educational systems per se linking the to an educational system, such as the tutor-web, would make the content available to a larger audience and save work on creating material within the educational system. Examples of existing content providers are Khan Academy and Connexions. A number of academic institutions have also made educational material available, including MIT and Stanford Engineering Everywhere. Finally several companies have emerged providing free online courses such as Coursera, edX and Udacity.

Recently AIWBES have become a popular subject of research. Several systems have been developed, many of which focus on a specific subject, often within computer science. Examples of AIWBES systems used in computer science education are SQL-Tutor [10], QuizGuide [11, 12] and Flip [13] which includes an interesting way of allocatingquiz questions to students.

## 3   THE TUTOR-WEB

The tutor-web system has been designed by considering four major requirements:

1. be open source and generic, not topic-specific
2. provide a wide range of open content through a web browser
3. use intelligent methods for quiz question allocation, amenable for research
4. function as a LCMS

As presented in Section 2, a substantial number of systems is available. However, none of them fulfills the four criteria stated above, to our knowledge.

A short introduction to the tutor-web is given in the following section. For a more detailed description see [14].

### 3.1   The organization of educational material

The teaching material in the system is organized into a tree (Fig. 1). Slides are grouped together to form lectures which are grouped into tutorials. A tutorial can belong to more than one course and should be built up around a single theme. For example, a tutorial on simple linear regression could both be a part of a general course on regression and an introductory statistics course. Quiz questions are linked to every lecture (Section 3.3).

Four types of users are defined in the tutor-web: regular users, students, teachers and managers. All tutor-web users can view and download teaching material but in order to answer quiz questions the user needs to become a tutor-web student by filling out a simple form. The student also needs to agree to that grades he or she earns in the system are recorded into a database and maybe used anonymously for research purposes. A tutor-web teacher can add and edit tutorials, lectures, slides and quizzes while managers can in addition add and edit departments and courses and give teacher rights.

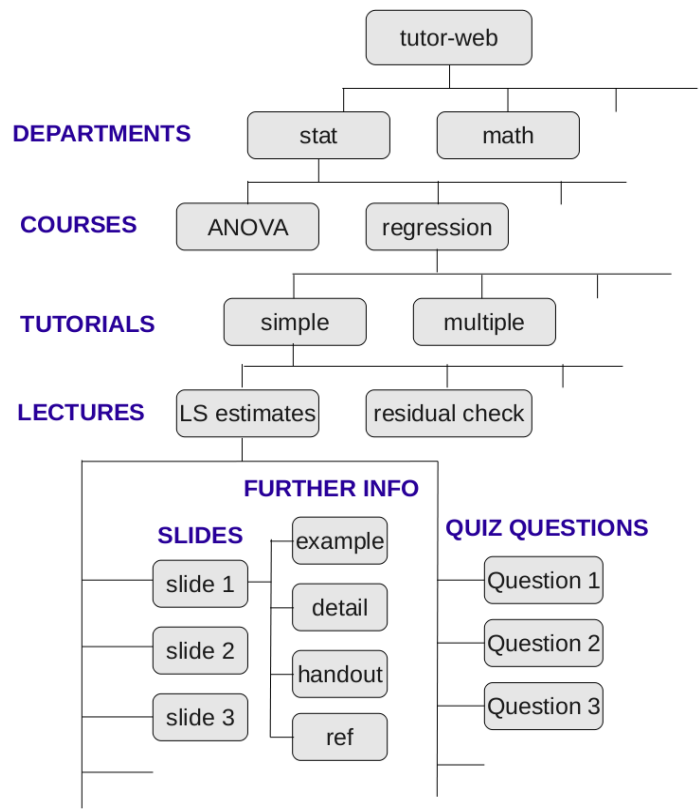

Figure 1: The structure of the tutor-web

## 3.2 Adding and viewing material

Teaching material can be added easily to the system through a web browser. A tutor-web teacher can built up a new tutorial with lectures including collection of slides. The slide is the core unit of the tutor-web. The other units are simply collection of information from the slides. A slide has a title, some text and/or figure(s). The format of the text can be LaTeX [15], plain text, Structured text [16] or HTML. The figure(s) can be uploaded files (png, gif or jpeg) or they can be rendered from a text based image format (R [17] or Gnuplot [18]). The teacher can choose to link some additional material to a slide such as examples, more details and/or references.

There are three different ways for a tutor-web user to view the teaching material (Fig. 2). The user can enter a lecture and browse through the material slide by slide. Links are provided to additional material attached to the slide (examples, references, ...) if any. The user can also download a PDF document including all the slides belonging to the lecture. The slides do not include the additional material linked to them (if any). The slides, in PDF format, are made with the LaTeX package Beamer [19] and should be ready to be used in a classroom. The third way of viewing teaching material in the tutor-web is on the tutorial level. Users can download a PDF document including all lectures belonging to that tutorial. A figure of the slides is provided in the document along with all additional material attached to them providing a handout including all relevant information. In a fully developed tutorial this corresponds to a complete textbook.

## 3.3 Quiz questions and grading

Quiz questions (items) are grouped together so they correspond to material within a lecture. The primary purpose of the tutor-web quizzes is to increase learning rather than mere evaluation. Students should therefore be allowed to answer quizzes at their own leisure. There is no built-in penalty for late answers, but an instructor may of course decide to specify a date at which a grade is extracted from the system.

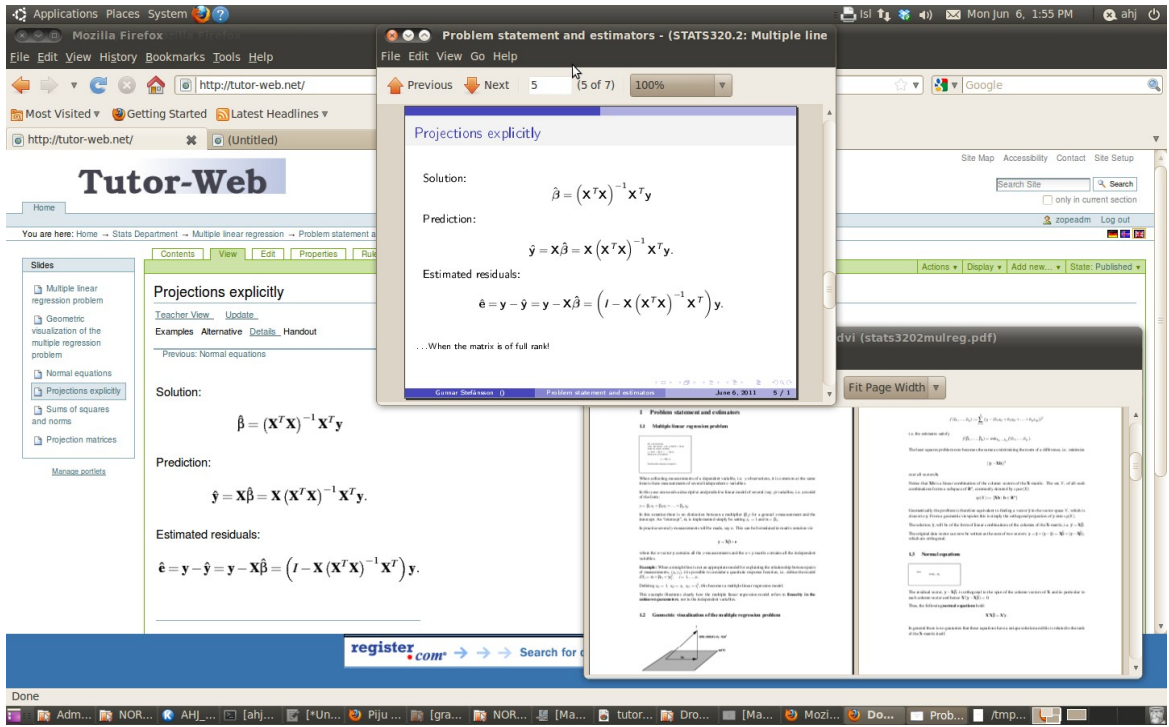

Figure 2: Different views into the database of teaching material in the tutor-web.

Questions and answers can be added to the system through a browser or be uploaded from a file. A quiz question can have as many answers as desired and there is an option to randomize the order of the answers. The format of the text can be LaTeX, plain text or Structured text. The system also allows the use of the statistical package R [17] when a question is generated. This allows the generation of similar but not identical data sets and graphs for students to analyse or interpret.

An algorithm us used to allocate appropriate questions to students where a probability mass function that depends on the grade of the student is used. The idea is that the system collects information on how often a question is allocated and how often it is answered correctly. For this purpose the difficulty of the question is simply calculated as

$$1 - \frac{\text{number of correct answers}}{\text{number of times question is answered}}.$$

The questions are then ranked according to their difficulty, from the easiest question to the most difficult one. A student with a low grade should have higher probability of getting the first questions after the ranking while a student with a high grade should have higher probabilities of getting the last questions. The mass of the probability function should therefore move towards the difficult questions as the grade goes up. A probability function in a lecture with 100 questions is shown in Fig. 3.

Currently a student gets one point for answering a question correctly and -1/2 for a wrong answer. Since the purpose of the quizzes is to allow the students to learn and thus improve the grade, only the last eight questions the student answers are used when the grade is calculated for each lecture. The student can track the grade with a press of a button, allowing each individual to monitor personal progress.

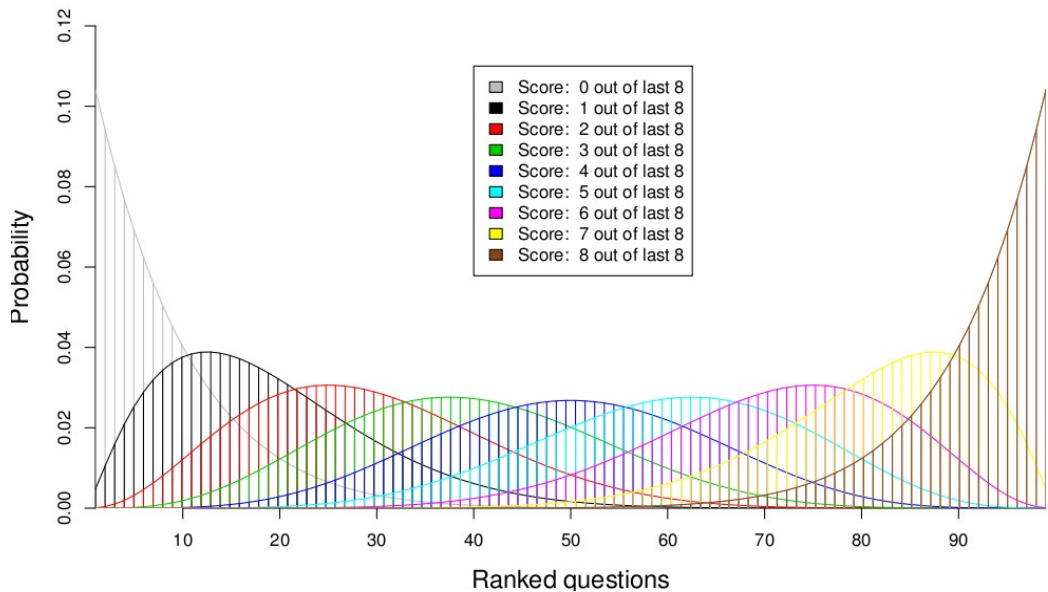

Fig. 3: Probability mass function for item allocations with 100 questions. The nine different colours represent students that have answered none up to all the last eight questions correctly.

## 4  THE RANDOMIZED TRIAL

An experiment was conducted to assess potential difference in learning between students using the tutor-web and students doing regular homework. The experiment was a randomized crossover experiment (Fig. 4). The student group consisted of students that had taken a course in calculus the semester before (strong mathematical background) and students that had not had any math lessons the last year, at least (weak mathematical background).

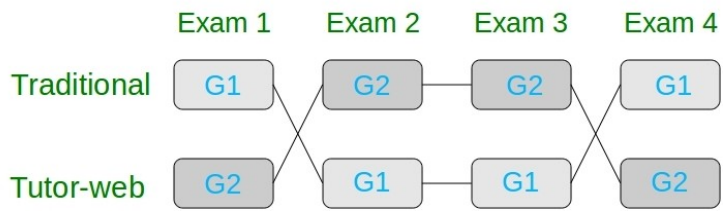

Figure 4: The design of the experiment.

The students were randomly split into two groups. In the first part of the experiment half of the students worked on quizzes in the tutor-web as homework while the other half handed in a written assignment. Shortly after the students handed in their homework they took an unexpected exam in class. In the next part of the experiment the groups were crossed, that is, the students that worked in the tutor-web before were told to hand in a written assignment and vice versa and again the students were tested. This was repeated two more times with 184 students taking at least one exam. The four subjects covered in the assignments and exams; discrete distributions, continuous distributions, hypotheses concerning means and contingency tables.

In order to test if there was any difference in learning in the groups, it was decided to fit the following analysis of variance model with the exam scores as the response variable:

$$y_{ijkl} = \beta_{1i} + \beta_{2j} + \beta_{3ij} + \beta_{4k} + \beta_{5l} + \epsilon_{ijkl}$$

where $β_1$ is the treatment effect, traditional/tutor-web (i = 1,2), $β_2$ is the math experience effect (j = 1,2), $β_3$ is the interaction between the treatment effect and the math experience effect, $β_4$ is the exam effect ($k = 1,2,3,4$) and $β_5$ is the student effect ($l = 1,2,...,183$).

The *lm* function in R [17] was used for the estimation. The results are shown in Table 1.

Table 1: Results of the initial ANOVA.

| Factor | Degrees of freedom | Sums of squares | F | p-value |
|---|---|---|---|---|
| Treatment | 1 | 2.7 | 1.01 | 0.32 |
| Math experience | 1 | 282.55 | 105.87 | $<2.2*10^{-16}$ |
| Interaction term | 1 | 0.02 | 0.01 | 0.93 |
| Exam | 3 | 421.94 | 52.7 | $<2.2*10^{-16}$ |
| Student | 182 | 1321.54 | 2.72 | $2.14*10^{-15}$ |

The interaction term between the treatment effect (tutor-web/traditional) and math experience effect was found to be insignificant (p = 0.93) and therefore removed from the model. The model was fitted again and the treatment effect was found to be insignificant (p = 0.63) and therefore removed. After removing the insignificant terms the model becomes

$$y_{ijkl} = β_{2j} + β_{3ij} + β_{5l} + ϵ_{ijkl}$$

The estimated confidence interval for the difference between the groups doing regular homework and working within the tutor-web was found to be $-0.23 < β_1 < 0.38$, before removed from the model. This indicates that there is no difference in mean test scores between students working within the tutor-web and students doing regular homework. This implies that time and money spent on correcting assignments can be saved by using the tutor-web without reducing the quality of the teaching.

## 5   CONCLUSIONS AND FUTURE WORK

A learning environment, the tutor-web has been developed and used for educational research. Educational material in mathematics and statistics can be found in the system which is open to everybody having access to the Web. In addition to well structured educational material the system offers quizzes with the purpose of increase learning rather than evaluation.

A randomized cross-over experiment was conducted to asses the difference in learning between students using the tutor-web and students doing regular homework. The difference in mean exam scores between the groups was not significant. This indicates that time and money can be saved by using the tutor-web as homework in stead of written assignments to some extent. The tutor-web is an ongoing research project. Similar experiments as described above will be conducted in 2014 while the system will be developed further.

## 6   ACKNOWLEDGEMENT

The tutor-web project has been supported by the Marine Research Institute of Iceland, the United Nations University and the University of Iceland.